\documentclass[11pt,preprint,showpacs,preprintnumbers,amsmath,amssymb]{revtex4}

\usepackage{graphicx}
\usepackage{dcolumn}
\usepackage{bm}
\usepackage{multirow}
\usepackage{graphicx}
\usepackage{amssymb}
\usepackage{float}
\usepackage{CJK}
\usepackage{caption}
\usepackage{subfigure}
\usepackage{relsize}
\usepackage{diagbox}
\usepackage{epstopdf}

\begin{document}

\title{$B_{s1}(5778)$ as a $B^*\bar{K}$ molecule in the Bethe-Salpeter equation approach}

\author{Zhen-Yang Wang \footnote{e-mail: wangzhenyang@nbu.edu.cn}}
\affiliation{\scriptsize{Physics Department, Ningbo University, Zhejiang 315211, China}}

\author{Jing-Juan Qi \footnote{e-mail: qijj@mail.bnu.edu.cn}}
\affiliation{\scriptsize{Junior College, Zhejiang Wanli University, Zhejiang 315101, China}}

\author{Qi-Xin Yu \footnote{e-mail: yuqx@mail.bnu.edu.cn}}
\affiliation{\scriptsize{Institute for Experimental Physics, Department of Physics, University of Hamburg, Luruper Chaussee 149, D-22761 Hamburg, Germany}}

\author{Xin-Heng Guo \footnote{Corresponding author, e-mail: xhguo@bnu.edu.cn}}
\affiliation{\scriptsize{College of Nuclear Science and Technology, Beijing Normal University, Beijing 100875, China}}

\date{\today}

\begin{abstract}
We interpret the $B_{s1}(5778)$ as an $S$-wave $B^\ast\bar{K}$ molecular state in the Bethe-Salpeter equation approach.  In the ladder and instantaneous approximations, and with the kernel containing one-particle-exchange diagrams and introducing three different form factors (monopole, dipole, and exponential form factors) in the vertex, we find the bound state exists. We also study the decay widths of the decay $B_{s1}(5778)\rightarrow B_s^\ast\pi$ and the radiative decays $B_{s1}(5778)\rightarrow B_s\gamma$ and $B_{s1}(5778)\rightarrow B_s^{\ast}\gamma$, which will be instructive for the forthcoming experiments.
\end{abstract}


\maketitle
\section{Introduction}
In the past decade lots of exotic hadrons were undoubtedly observed in experiments, like the $X$, $Y$, $Z$, and $P_c$ states \cite{Tanabashi:2018oca} (see detailed reviews in Refs. \cite{Chen:2016qju,Esposito:2016noz,Lebed:2016hpi,Dong:2017gaw,Ali:2017jda,Olsen:2017bmm,Guo:2017jvc}). These exotic hadrons do not meet the expectations of the quark model, therefore, explanations of their internal structures have been a very important topic. Among various explanations of the possible internal structures of these exotic hadrons, the hadronic molecule is a popular one. One of the main reasons to treat these observed exotic hadrons as molecules is that their masses are close to the thresholds of corresponding hadron pairs.

In the open charm orbitally excited mesons, there are some candidates of hadronic molecules. Particular interest has been paid to the positive-parity charm-strange mesons $D_{s0}^\ast(2317)$ and $D_{s1}(2460)$ observed in 2003 by $BABAR$ \cite{Aubert:2003fg} and CLEO \cite{Besson:2003cp} Collaborations. The masses of $D_{s0}^\ast(2317)$ and $D_{s1}(2460)$ are about 160 MeV and 70 MeV below the predicted $0^+$ and $1^+$ charm-strange mesons by the quark model \cite{Godfrey:1985xj,DiPierro:2001dwf}, respectively. Since their masses are below the $DK$ and $D^\ast K$ thresholds by about 45 MeV, they are natural candidates for hadronic molecules. Up to now, the inner structure, strong and radiative decays of $D_{s0}^\ast(2317)$ and $D_{s1}(2460)$ in the molecule picture have been studied in different approaches \cite{Barnes:2003dj,Chen:2004dy,Guo:2006fu,Guo:2006rp,Gamermann:2006nm,Xie:2010zza,Feng:2012zze,Faessler:2007gv,Faessler:2007us,Mohler:2013rwa,Xiao:2016hoa} such as the quark model, the effective Lagrangian approach, the Bethe-Salpeter approach, and lattice QCD.

So far, the $b$-partners of $D_{s0}^\ast(2317)$ and $D_{s1}(2460)$, $B_{s0}^\ast$ and $B_{s1}$, have not been observed. However, there have been a lot of theoretical efforts to investigate the properties of the $B_{s0}$ and $B_{s1}^\ast$ states, e.g. the mass spectroscopy, strong decays and radiative decays in different models \cite{Lang:2015hza,Altenbuchinger:2013vwa,DiPierro:2001dwf,Ebert:2009ua,Sun:2014wea,Godfrey:2016nwn,Bardeen:2003kt,Lu:2016bbk,Gregory:2010gm,Cleven:2010aw,Guo:2006fu,Guo:2006rp,Kolomeitsev:2003ac,Colangelo:2012xi,Cheng:2014bca,Faessler:2008vc,Zhong:2008kd} assumed them as excited states ($q\bar{q}$), four-quark states or molecular states. There are large discrepancies between the existing theoretical results of different models. Hence, more careful studies are needed, especially in the relativistic models, because the mass of the light quark ($K$ meson) is rather small compared to $b$ quark ($B$/$B^\ast$ meson), and relativistic corrections are expected to be large.

The $B_{s0}^\ast$ state as a $B\bar{K}$ bound state was studied in our previous work \cite{Feng:2012zze}. In this paper, we will focus on the $B_{s1}$ in the Bethe-Salpeter (BS) equation approach which is a relativistic method. We assume that the $B_{s1}$ is an $S$-wave $B^\ast\bar{K}$ bound state taking the isospin, spin, and parity quantum numbers of the $B_{s1}$ as $I(J^P)=0(1^+)$, and the mass $m_{B_{s1}}=5778$ MeV (the central value predicted in Ref. \cite{Guo:2006rp}). One purpose of the present paper is to investigate whether the bound state of the $B^\ast\bar{K}$ system exist. The other one is to study the isospin-violating decay $B_{s1}\rightarrow B_s^\ast\pi^0$ and the radiative decays $B_{s1}\rightarrow B_s\gamma$ and $B_{s1}\rightarrow B_s^\ast\gamma$ if $B_{s1}$ could be the $B^\ast\bar{K}$ molecule.

In the rest of the manuscript we proceed as follows. In Sec. \ref{sect-BS}, we establish the BS equation for the bound state of a vector meson ($B^\ast$) and a pseudoscalar meson ($K$). Then we discuss the interaction kernel of the BS equation and calculate corresponding numerical results of the Lorentz scalar functions in the normalized BS wave function in Sec. \ref{Nor-BS} . In Sec. \ref{decay}, the decay widths of the $B^\ast\bar{K}$ bound state to $B_s\pi$, $B_s^\ast\gamma$, and $B_s^\ast\gamma$ final states are calculated. In Sec. \ref{sum} we present a summary of our results.

\section{the bethe-salpeter formalism for $B^\ast\bar{K}$ system}
\label{sect-BS}
In this section we discuss the formalism for the study of the $B_{s1}(5778)$ as a $B^\ast\bar{K}$ hadronic molecule with the BS approach. Since the isospin quantum number of $B_{s1}(5778)$ is 0, for the $B^\ast\bar{K}$ system, the flavor wave function of the isoscalar bound state is $|P\rangle_{0,0}=\frac{1}{\sqrt{2}}|B^{\ast+}K^-\rangle-\frac{1}{\sqrt{2}}|B^{\ast0}\bar{K}^0\rangle$, where the subscript $(0,0)$ refers to the isospin and its third component.

The BS wave function for the bound state $|P\rangle$ of a vector meson ($B^\ast$) and a pseudoscalar meson ($\bar{K}$) is defined as the following:
\begin{equation}
  \chi^{\mu}_P\left(x_1,x_2,P\right) = \langle0|TB^{\ast\mu}(x_1)\bar{K}(x_2)|P\rangle,
\end{equation}
where $B^{\ast\mu}(x_1)$ and $\bar{K}(x_2)$ are the field operators of the vector and pseudoscalar mesons at space coordinates $x_1$ and $x_2$, respectively, $P$ denotes the total momentum of the bound state with mass $M$ and velocity $v$.

\begin{figure}[ht]
\centering
    \rotatebox{0}{\includegraphics*[width=0.35\textwidth]{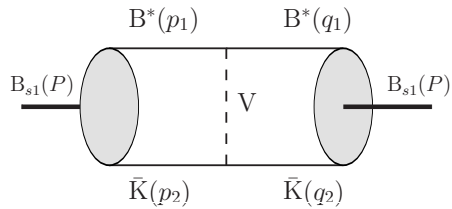}}
    \caption{One-particle exchange diagrams induced by a vector mesons $\rho$ and $\omega$.}
  \label{bound-state}
\end{figure}

The BS equation for the bound state can be written in the following form:
\begin{equation}\label{BS-equation}
  \chi_{P}^{\mu}(p)=S_{B^\ast}^{\mu\nu}(p_1)\int\frac{d^4q}{(2\pi)^4}K_{\nu\sigma}(P,p,q)\chi_{P}^\sigma(q)S_K(p_2),
\end{equation}
where $S_{B^\ast}^{\mu\nu}(p_1)$ and $S_K(p_2)$ are the propagators of the vector and pseudoscalar mesons, $B^\ast$ and $K$, respectively. $K_{\nu\sigma}(P,p,q)$ is the kernel which contains two-particle-irreducible diagrams ($p$ and $q$ are the relative momenta of the initial and final constituent particles, respectively). The kernel will be calculated based on the Feynman diagrams shown in Fig. \ref{bound-state} using the chiral Lagrangian, at the tree level and in the $t$-channel.

For convenience, we define $p_l (=p\cdot v)$ and $p_t^\mu(=p^\mu- p_lv^\mu)$ to be the longitudinal and transverse  projections of the relative momentum ($p=\lambda_2 p_1-\lambda_1 p_2$ with $\lambda_i=m_i/(m_1+m_2)$, $m_i(i=1,2)$ being the mass of the $i$-th constituent particle) along the bound state momentum ($P$). Then the propagator of the $B^\ast$ meson in the heavy quark limit can be expressed as follows:
\begin{equation}\label{vector-propagator}
  S^{\mu\nu}_{B^\ast}(\lambda_1P+p)=\frac{-i\left(g^{\mu\nu}-v^\mu v^\nu\right)}{2\omega_1\left(\lambda_1M+p_l-\omega_1+i\epsilon\right)}.
\end{equation}
The propagator of the $\bar{K}$ meson is
\begin{equation}\label{pseudoscalar-propagator}
  S_K(\lambda_2P-p)=\frac{i}{\left(\lambda_2M-p_l\right)^2-\omega_2^2+i\epsilon}.
\end{equation}
In Eqs. (\ref{vector-propagator}) and \ref{pseudoscalar-propagator}) $\omega_{1(2)} = \sqrt{m_{1(2)}^2+p_t^2}$ (we have defined $p_t^2=-p_t\cdot p_t$).

To describe the interaction between the heavy vector meson and the light pseudoscalar meson, we employ the following chiral Lagrangian:
\begin{equation}\label{BS-lagran}
\begin{split}
  \mathcal{L}_{KKV}&=ig_{KK\rho}(K_b\partial_\mu K_a^\dag-\partial_\mu K_b K_a^\dag)V_{ba}^\mu,\\
  \mathcal{L}_{B^*B^*V}&=ig_{B^*B^*V}(B^{*\nu}_b\stackrel{\leftrightarrow}{\partial_{\mu}}B^{*\dag}_{\nu,a})V^\mu_{ba}+if_{B^*B^*V}(B^{*}_{\mu, b} B^{*\dag}_{\nu,a}-B^{*\dag}_{\mu,a} B^*_{\nu,b})(\partial^\mu V^\nu-\partial^\nu V^\mu)_{ba},
  \end{split}
\end{equation}
where $A\stackrel{\leftrightarrow}{\partial_{\mu}}B\equiv A\partial B-\partial A B$, and $V$ is the matrix of the nonet vector meson,
\begin{equation}\label{eq:vector}
V=\left( \begin{array}{ccc}\frac{\rho^0}{\sqrt{2}}+\frac{\omega}{\sqrt{2}}&\rho^+&K^{*+}\\
\rho^-&-\frac{\rho^0}{\sqrt{2}}+\frac{\omega}{\sqrt{2}}&K^{*0}\\
K^{*-}&\bar{K}^{*0}&\phi\\
\end{array} \right),
\end{equation}
and the coupling constants are given as $g_{KK\rho}=g_{KK\omega}=g_{V}/2$, $g_{B^\ast B^\ast V}=-\frac{1}{\sqrt{2}}\beta g_V$, and $f_{B^\ast B^\ast V}=-\sqrt{2}\lambda g_V m_{B^\ast}$ with the parameter $g_V =5.8$ being determined by the Kawarabayashi-Suzuki-Riazuddin-Fayyazuddin relations \cite{Ding:2008gr}, the parameter $\beta$ is estimated to be about 0.9, and the parameter $\lambda$ is obtained by light-cone sum rule and lattice QCD, $\lambda=0.56 $ ${\rm GeV}^{-1}$ \cite{Isola:2003fh}.

The kernel of the BS equation at the tree level and in the $t$-channel with the so-called ladder approximation can be obtained as following:
\begin{equation}\label{kernel}
\begin{split}
  K^{\mu\nu}&(p_1,p_2;q_2,q_1)
  =-c_I\Big\{g_{B^*B^*V}g_{KKV}g_{\mu\nu}(p_1+q_1)_\sigma(p_2+q_2)_\alpha\Delta^{\alpha\sigma}(k,m_V)\\
                             &+f_{B^*B^*V}g_{KKV}(p_2+q_2)_\alpha\left[k_\mu\Delta^\alpha_\nu(k,m_V)-k_\nu\Delta^\alpha_\mu(k,m_V)\right]\Big\}(2\pi)^4\delta^4(q_1+q_2-p_1-p_2),
\end{split}
\end{equation}
where $c_I$ is the isospin coefficient: $c_I$ = 3 for $\rho$ exchange diagram while $c_I$ = 1 for $\omega$ exchange diagram, $\Delta_{\mu\nu}(k,m_V )$ (V = $\rho,\omega$) denotes the massive vector meson propagator which has the following form:
\begin{equation}
  \Delta_{\mu\nu}(k,m_V )=\frac{-i}{k^2-m_V^2}\left(g_{\mu\nu}-\frac{k_\mu k_\nu}{m_V^2}\right),
\end{equation}
where $m_V$ is the mass of the exchanged meson and $k$ is its momentum.

In order to describe finite size effect of the interacting hadrons at the vertex, we introduce a form factor $\mathcal{F}(k)$ at each vertex. Generally, the form factor could have the monopole, dipole, and exponential forms as shown below, respectively,
\begin{equation}\label{form-factor}
  \begin{split}
  \mathcal{F}_M(k)&=\frac{\Lambda^2-m_V^2}{\Lambda^2-k^2},\\
  \mathcal{F}_D(k)&=\frac{(\Lambda^2-m_V^2)^2}{(\Lambda^2-k^2)^2},\\
  \mathcal{F}_E(k)&=e^{(k^2-m^2)/\Lambda^2},\\
  \end{split}
\end{equation}
where $\Lambda$ is a phenomenological cutoff which will be adjusted in a reasonable range while solving the BS equation.

In general, for a vector meson ($B^\ast$) and a pseudoscalar meson ($K$) bound state, the BS wave function $\chi_P^\mu(p)$ has the following form:
\begin{equation}
\chi_P^\mu(p)=f_0(p)p^\mu+f_1(p)P^\mu+f_2(p)\epsilon^\mu+f_3(p)\varepsilon^{\mu\nu\alpha\beta}p_\alpha P_\beta\epsilon_\nu,
\end{equation}
where $f_i(p)$ $(i = 0,1,2,3)$ are Lorentz-scalar functions. After considering the constraints imposed by parity and Lorentz transformations, it is easy to prove that $\chi_P^\mu(p)$ can be simplified as
\begin{equation}\label{BS-wave-function}
  \chi_P^\mu(p)=f(p)\varepsilon^{\mu\nu\alpha\beta}p_\alpha P_\beta\epsilon_\nu,
\end{equation}
where the function $f(p)$ contains all the dynamics and $\epsilon^\mu$ represents the polarization vector of the bound state.

In the following calculation, we will use the covariant instantaneous approximation ($p_l=q_l$) in which the energy exchanged between the constituent particles of the binding system is neglected. Since in the heavy quark limit the heavy meson ($B^\ast$) is almost on-shell and the binding of the constituent particles is weak, it is appropriate to use this approximation so that the longitudinal momentum of the exchanged meson is put to zero in the kernel \cite{Dai:1993kt,Guo:1996jj}.

Using the covariant instantaneous approximation and substituting Eqs.(\ref{vector-propagator}), (\ref{pseudoscalar-propagator}), and (\ref{kernel}) into Eq.(\ref{BS-equation}), we have
\begin{equation}\label{4-p-BS-equation}
  \begin{split}
  f(p)=&\int\frac{d^4q}{(2\pi)^4}\frac{i}{6\omega_1(\lambda_1M+p_l-\omega_1+i\epsilon)[(\lambda_2M-p_l)^2-\omega_2^2+i\epsilon][-(p_t-q_t)^2-m_V^2]}\\
       &\Big\{3g_{B^*B^*V}g_{KKV}\left[4(\lambda_1M+p_l)(\lambda_2M-p_l)+(p_t+q_t)^2+(p_t^2-q_t^2)^2/m_V^2\right]\\
       &+f_{B^*B^*V}g_{KKV}\omega_2(p_t\cdot q_t-q_t^2)/(\lambda_2M-\omega_2))\Big\}F^2(k_t)f(q).
  \end{split}
\end{equation}
where $k_t=p_t-q_t$ is the momentum of the exchanged meson in the covariant instantaneous approximation.

In Eq. (\ref{4-p-BS-equation}) there are poles in the plane $p_l$ at $-\lambda_1 M+\omega_1-i\epsilon$, $\lambda_2 M+\omega_2-i\epsilon$ and $\lambda_2 M-\omega_2+i\epsilon$. By choosing the appropriate contour, we integrate over $p_l$ on both sides of Eq. (\ref{4-p-BS-equation}) in the rest frame of the bound state, we obtain the following equation
\begin{equation}\label{3-p-BS-equation}
\begin{split}
\tilde{f}(p_t)=&\int\frac{dq_t^3}{(2\pi)^3}\frac{1}{12\omega_1\omega_2(M-\omega_1-\omega_2)[-(p_t-q_t)^2-m_V^2]}\\
&\times\Big\{3g_{B^*B^*V}g_{KKV}[4\omega_2(M-\omega_2)+(p_t+q_t)^2+(p_t^2-q_t^2)^2/m_V^2]\\
&+2f_{B^*B^*V}g_{KKV}\omega_2(p_t\cdot q_t-q_t^2)/(\lambda_2M-\omega_2)\Big\}\tilde{f}(q_t),
\end{split}
\end{equation}
where $\tilde{f}(p_t)\equiv\int dp_l f(p)$.

Now, we can solve the BS equation numerically and study whether the $S$-wave $B^\ast \bar{K}$ bound state exists or not. It can be seen from Eq. (\ref{3-p-BS-equation}) that there is only one free parameter in our model, the cutoff $\Lambda$, which enters through various phenomenological form factors in Eq. (\ref{form-factor}).  It contains the information about the extended interaction due to the structures of hadrons. The value of $\Lambda$ is of order 1 GeV which is the typical scale of nonperturbative QCD interaction. In this work, we shall treat $\Lambda$ as a parameter and vary it in a much wider range 0.8-4.8 GeV \cite{Wang:2017dcq,Wang:2018jaj} when the binding energy $E_b$ (which is defined as $E_b=M-m_1-m_2$) is in the region 0 to -100 MeV to see if the BS equation has solutions.

To find out the possible molecular bound states, one only needs to solve the homogeneous Bethe-Salpeter equation.  One numerical solution of the homogeneous Bethe-Salpeter equation corresponds to a possible bound state.  The integration region in each integral is discretized into $n$ pieces, with $n$ being sufficiently large. In this way, the integral equation is converted into an $n\times n$ nmatrix equation, and the scalar wave function will now be regarded as an $n$-dimensional vector. Then, the integral equation can be illustrated as $\tilde{f}^{(n)}(|p_t|)=A^{(n\times n)}(|p_t|,|q_t|)\tilde{f}^{(n)}(|q_t|)$, where $\tilde{f}^{(n)}(|p_t|) (\tilde{f}^{(n)}(|q_t|))$ is an $n$-dimensional vector, and $A^{(n\times n)}(|p_t|,|q_t|)$ is an $n \times n$ matrix, which corresponds to the matrix labeled by $p_t$ and $q_t$ in each integral equation. Generally, $|p_t|$ ($|q_t|$) varies from 0 to $+\infty$. Here, $|p_t|$ ($|q_t|$) is transformed into a new variable $t$ that varies from $-1$ to 1 based on the Gaussian integration method,
\begin{equation}
|p_t|=\mu+w\log\left[1+y\frac{1+t}{1-t}\right],
\end{equation}
where $\mu$ is a parameter introduced to avoid divergence in numerical calculations, $w$ and $y$ are parameters used in controlling the slope of wave functions and finding the proper solutions
for these functions. Then one can obtain the numerical results of the Bethe-Salpeter wave functions by requiring the eigenvalue of the eigenvalue equation to be 1.

In our calculation, we choose to work in the rest frame of the bound state in which $P=(M,0)$. We take the averaged masses of the mesons from the PDG, $m_{B^\ast}=5324.65$ MeV, $m_{K}=494.98$ MeV, $m_\rho=775.26$ MeV, and $m_\omega=782.65$ MeV. After searching for possible solutions in the isoscalar channel of the $B^\ast \bar{K}$ system, we find the bound state exists. We list some values of $E_b$ and the corresponding $\Lambda$ for the three different form factors in Table \ref{Eb-Lambda}.

\begin{table}[tb]
\renewcommand{\arraystretch}{1.2}
\centering
\caption{
Values of $E_b$ and corresponding cutoff $\Lambda$, $\Lambda_M$, $\Lambda_D$, and $\Lambda_E$ for the monopole, dipole, and exponential form factors, respectively.}
\begin{tabular*}{\textwidth}{@{\extracolsep{\fill}}ccccccccccc}
\hline
\hline
$E_b$(MeV)   &  -10  &  -20  &  -30  &  -40  &  -50  &  -60  &  -70  &  -80  &  -90  &  -100 \\
\hline
$\Lambda_M$(MeV)  &  1350  &  1428  &  1485  &  1531  &  1571  &  1608  &  1641  &  1672  &  1701  &  1728 \\
$\Lambda_D$(MeV)  &  1897  &  2025  &  2118  &  2194  &  2261  &  2320  &  2375  &  2425  &  2473  &  2518 \\
$\Lambda_E$(MeV)  &  1340  &  1443  &  1517  &  1578  &  1632  &  1680  &  1723  &  1764  &  1803  &  1839 \\
\hline
\hline
\end{tabular*}\label{Eb-Lambda}
\end{table}

\section{The Normalization Condition of the Bethe-Salpeter wave function}
\label{Nor-BS}
To find out whether the bound state of the $B^\ast\bar{K}$ system exists or not, one only needs to solve the homogeneous BS equation. However, when we want to calculate physical quantities such as the decay width we have to face the problem of the normalization of the BS wave function. In the following we will discuss the normalization of the BS wave function $\chi^\mu_P(p)$.

In the heavy quark limit, the normalization of the BS wave function of the $B^\ast\bar{K}$ system can be written as \cite{Guo:2007qu}
\begin{equation}
i\int\frac{d^4pd^4q}{(2\pi)^8}\bar{\chi}^\mu_P(p)\frac{\partial}{\partial P_0}[I_{P\mu\nu}(p,q)]\chi^\nu(q)=1,
\end{equation}
where $I_{P\mu\nu}(p,q)=(2\pi)^4\delta^4(p-q)S^{-1}_{\mu\nu}(p_1)S^{-1}(p_2)$.

In the rest frame, the normalization condition can be written in the following form:
\begin{equation}
\begin{split}
-i\int\frac{d^4p}{(2\pi)^4}\Big\{&4M^2p_t^2\big[\lambda_1^2(6\lambda_2^2M^2-6\lambda_2Mp_l+p_l^2-\omega_2^2)\\
&+2\lambda_1\lambda_2p_l(3\lambda_2M-2p_l)+\lambda_2^2(p_l^2-\omega_1^2)\big]f^2(q)=1.
\end{split}
\end{equation}
From Eqs. (\ref{4-p-BS-equation}) and (\ref{3-p-BS-equation}), we obtain
\begin{equation}
  \begin{split}
  f(p)=&\frac{i\omega_1\omega_2(M-\omega_1-\omega_2)}{\pi(\lambda_1M+p_l-\omega_1+i\epsilon)(\lambda_2M-p_l+\omega_2-i\epsilon)(\lambda_2M-p_l-\omega_2+i\epsilon)}\tilde{f}(p_t).\\
  \end{split}
\end{equation}
Then, one can recast the normalization condition for the BS wave function into the form
\begin{equation}\label{3-Nor-Eq}
\begin{split}
&-\int\frac{d^3p_t}{8\pi^5}\frac{M^2p_t^2\omega_1}{\omega_2^2(M-\omega_1-\omega_2)^2}\Big\{\lambda_2^2(p_t^2-\omega_1^2)(\lambda_2M-\omega_1-3\omega_2)+\lambda_1^3(\lambda_2^2M^3-2M\omega_2^2)\\
&+\lambda_1\lambda_2[2\lambda_2^3M^3+\lambda_2M(p_t^2-\omega_1^2)-4\omega_2^2(\omega_1-\omega_2)-2\lambda_2^2M^2(\omega_1+3\omega_2)]\\
&+\lambda_1^2[3\lambda_2^3M^3-6\lambda_2M\omega_2^2+2\omega_2^2(\omega_1+\omega_2)-\lambda_2^2M^2(\omega_1+3\omega_2)]\Big\}\tilde{f}^2(p_t)=1.
\end{split}
\end{equation}

The wave function obtained in the previous section (which is calculated numerically from Eq.(\ref{3-p-BS-equation})) can be normalized by Eq. (\ref{3-Nor-Eq}).

\begin{figure}[htbp]
\centering
\subfigure[]{
\begin{minipage}[t]{0.3\linewidth}
\centering
\includegraphics[width=2.1in]{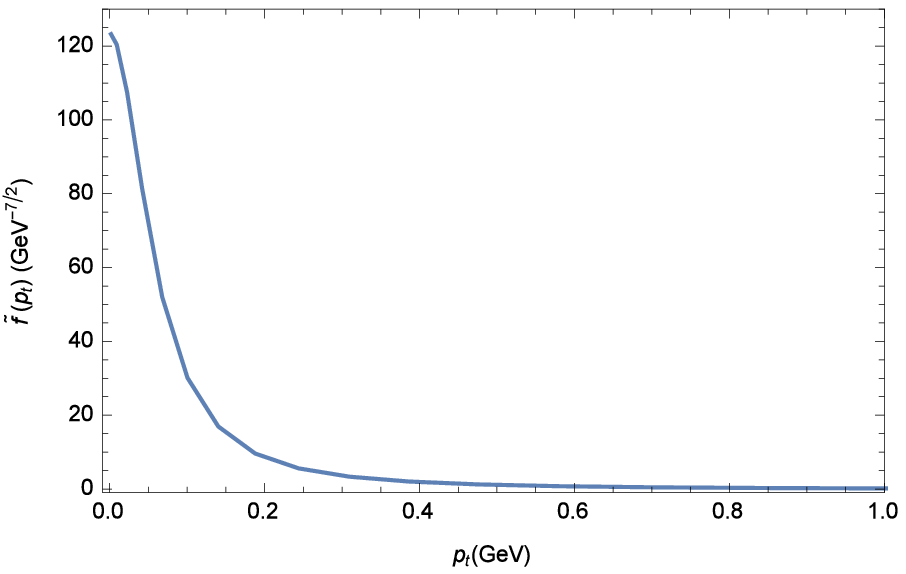}
\end{minipage}%
}%
\subfigure[]{
\begin{minipage}[t]{0.3\linewidth}
\centering
\includegraphics[width=2.1in]{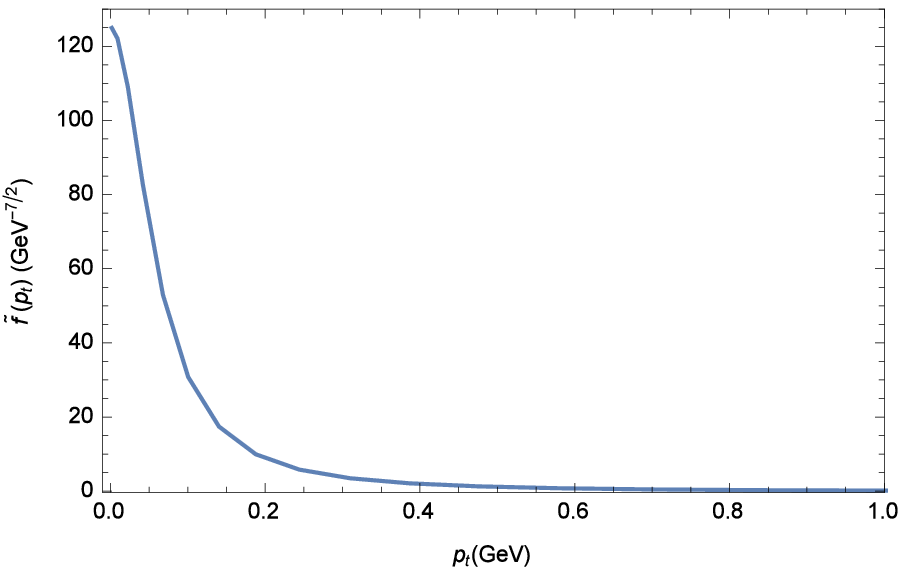}
\end{minipage}
}%
\subfigure[]{
\begin{minipage}[t]{0.3\linewidth}
\centering
\includegraphics[width=2.2in]{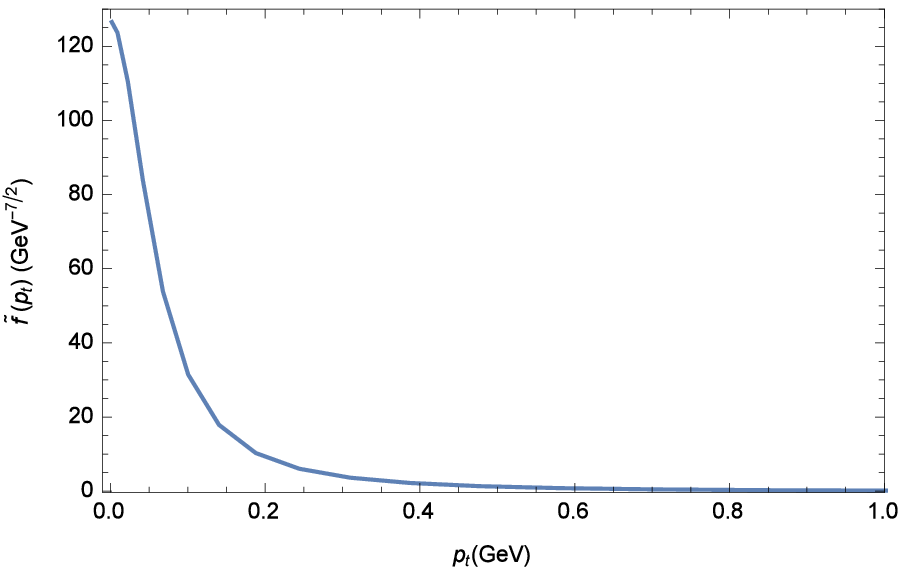}
\end{minipage}%
}%
\centering
\caption{Numerical results of the normalized Lorentz scalar function $\tilde{f}(p_t)$ for $B_{s1}(5778)$ in the $B^\ast\bar{K}$ molecular picture for (a) the monopole form factor, (b) the dipole form factor, and (c) the exponential form factor.}
\label{W-bound-state}
\end{figure}

In our case, the binding energy $E_b=M_{B_{s1}(5778)}-(M_{B^\ast}-M_K)\simeq 42.3$ MeV, where we have used the mass $B_{s1}(5778)$ as 5778 MeV. From our calculations, we find the $B^\ast\bar{K}$ system can be $B_{s1}(5778)$ state when the cutoff $\Lambda$ = 1541 MeV, 2210 MeV, and 1591 MeV for the monopole, dipole, and exponential form factors, respectively. The corresponding numerical results of the normalized Lorentz scalar function, $\tilde{f}(p_t)$, are given in Fig. \ref{W-bound-state} for the $B_{s1}(5778)$ state in the $B^\ast\bar{K}$ molecule picture for the monopole, dipole, and exponential form factors, respectively.

\section{Decays of $B_{s1}(5778)$}
\label{decay}
Besides investigating whether the bound state of the $B^\ast\bar{K}$ system can be $B_{s1}(5778)$ or not, we can also study other properties of this molecular bound state which can be measured in experiments. In the following we will study the decay widths of $B_{s1}(5778)$ decaying into $B_s^\ast\pi$, $B_s\gamma$ and $B_s^\ast\gamma$.

\subsection{The strong decay of $B_{s1}(5778)\rightarrow B_s^\ast\pi$}
In this subsection, we will calculate the decay width of the process $B_{s1}(5778)\rightarrow B_s^\ast\pi$ through exchangeing $B$, $B^\ast$, and $K^\ast$ mesons, Since this decay is a isospin-violating process, there exist two possible dynamical mechanisms: one is the so-called direct mechanism with $\pi^0$ emission from the $B\rightarrow B^\ast$ and $K\rightarrow K^\ast$ transitions, the other one is the indirect mechanism where a $\pi^0$ meson is produced via $\eta-\pi^0$ mixing. The direct and mixing mechanisms can be combined together in the form of an effective coupling of $\pi^0$ to the mesonic pairs $B B^\ast$ or $K K^\ast$ with modified flavour structure. Consequently, instead of the $\tau_3\pi^0$ coupling to $B B^\ast$ or $K K^\ast$ we have $\pi^0(\tau_3\cos\varepsilon+\kappa I\sin\varepsilon)$, where $\kappa=1/\sqrt{3}$ or $\sqrt{3}$ is the corresponding flavor-algebra factor for the $B B^\ast$ or $K K^\ast$ coupling, respectively. The $\eta-\pi^0$ mixing angle $\varepsilon$ is fixed as \cite{Faessler:2008vc}:
\begin{equation}
\tan2\varepsilon=\frac{\sqrt{3}}{2}\frac{m_d-m_u}{m_s-\hat{m}}\simeq0.02,\quad\quad \hat{m}=\frac12(m_u+m_d),
\end{equation}
where $m_u$, $m_d$, $m_s$ are the current quark masses.

As in Ref. \cite{Xiao:2016mho}, the effective Lagrangians relevant to the decay $B_{s1}(5778)\rightarrow B_s^\ast\pi$ are
\begin{equation}
  \begin{split}
\mathcal{L}_{B^{\ast}BV}=&-2f_{B^\ast BV}\varepsilon^{\mu\nu\alpha\beta}(B^\dag\stackrel{\leftrightarrow}{\partial_{\mu}}B^{\ast\nu}-B^{\ast\nu\dag}\stackrel{\leftrightarrow}{\partial_{\nu}}B)\partial_\alpha V_\beta\\
\mathcal{L}_{B^{(\ast)}B^{\ast}P}=&-ig_{B^\ast BP}(B^\dag\partial_\mu PB^{\ast\mu}-B^{\ast\mu\dag}\partial_\mu PB)+\frac12g_{B^\ast B^\ast P}\varepsilon^{\mu\nu\alpha\beta}B^{\ast\dag}_{\mu}\partial_\nu
                                   P\stackrel{\leftrightarrow}{\partial_{\alpha}}B^\ast_{\beta},\\
\mathcal{L}_{K^{\ast}K\pi}=&-ig_{K^{\ast}K\pi}K_\mu^{\ast\dag}\hat{\pi}\stackrel{\leftrightarrow}{\partial^{\mu}}K,\\
  \end{split}
\end{equation}
where $B^{(\ast)^\dag}=(B^{(\ast)-}, \bar{B}^{(\ast)0}, \bar{B}^{(\ast)0}_s)$and $P$ has the following form:
\begin{equation}
\label{eq:pseudo}
P=\left(
\begin{array}{ccc}\frac{\pi^0}{\sqrt{2}}+\frac{\eta}{\sqrt{6}}&\pi^+&K^+\\
\pi^-&-\frac{\pi^0}{\sqrt{2}}+\frac{\eta}{\sqrt{6}}&K^0\\
K^-&\bar{K}^0&-\frac{2\eta}{\sqrt{6}}\\
\end{array} \right),
\end{equation}
The effective Lagrangians for $B^\ast B^\ast V$ vertex and $V$ has been given in Eq. (\ref{BS-lagran}). The coupling constants of the bottom mesons to the light mesons could be evaluated with the aid of the heavy quark symmetry and the chiral symmetry. The coupling constants $g_{B^{(\ast)}B^\ast P}$ are related to a coupling constant $g$ by
\begin{equation}
g_{B^{\ast}B^\ast P}=\frac{2g}{f_\pi},\quad\quad g_{B^{\ast}B P}=\frac{2g}{f_\pi}\sqrt{m_{B^\ast}m_B},
\end{equation}
where $f_\pi$ = 132 MeV is the pion decay constant and $g$=0.44 is determined by the lattice QCD caculation \cite{Becirevic:2009yb}. The coupling constant $f_{B^\ast BV} =\lambda g_V/\sqrt{2}$ as shown in Refs. \cite{Lin:1999ad,Oh:2000qr}. We use $g_{K^\ast K\pi}=3.21$ as in Ref. \cite{Liu:2005jb}.

\begin{figure}[ht]
\centering
    \rotatebox{0}{\includegraphics*[width=0.50\textwidth]{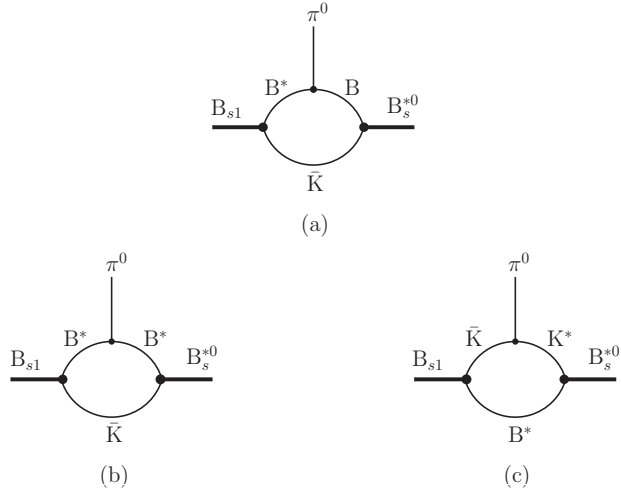}}
    \caption{Diagrams contributing to the $B_{s1}\rightarrow B_s^{*0}\pi^0$ decay.}
  \label{decay1}
\end{figure}

According to the above interactions, the decay diagrams $B_{s1}(5778)\rightarrow B_s^\ast\pi$ induced by $B$, $B^\ast$, and $K^\ast$ exchanges are shown in Fig. \ref{decay1} and the corresponding amplitudes are

\begin{equation}
\begin{split}
\mathcal{M}_{4a}&=g_{B^\ast B\pi}g_{B_s^\ast BK}\epsilon^\nu_{B_s^\ast}\int\frac{d^4p}{(2\pi)^4}p'_{1\mu}p_{2\nu}\Delta_B(k,m_B)\mathcal{F}^2(k)\chi_P^\mu(p),\\
\mathcal{M}_{4b}&=-\frac14g_{B^\ast B^\ast\pi}g_{B_s^\ast B^\ast K} \epsilon_{\alpha\rho\sigma\mu}\epsilon_{\beta\kappa\lambda\nu}\epsilon_{B_s^\ast}^\nu\int\frac{d^4p}{(2\pi)^4}p^{'\rho}_{1}p_2^\kappa(p_1-k)^\sigma(p'_2-k)^\lambda\Delta_{B^\ast}^{\alpha\beta}(k,m_{B^\ast})\mathcal{F}^2(k)\chi_{P}^\mu(p),\\
\mathcal{M}_{4c}&=g_{K^\ast K\pi}\epsilon^\nu_{B_s^\ast}\int\frac{d^4p}{(2\pi)^4}(p_2+p'_1)_\beta [g_{B^\ast_sB^\ast K^\ast}(p_1+p'_2)_\alpha g_{\mu\nu}+4f_{B_s^\ast B^\ast K^\ast}(k_\nu g_{\alpha\mu}-k_\mu g_{\alpha\nu})]\\
&\times\Delta_{K^\ast}^{\alpha\beta}(k,m_{K^\ast})\mathcal{F}^2(k)\chi_{P}^\mu(p).
\end{split}
\end{equation}
The total amplitude of $B_{s1}(5778)\rightarrow B_s^\ast\pi$ is then
\begin{equation}
\mathcal{M}^{tot}_{B_{s1}\rightarrow B_s^\ast\pi}=\mathcal{M}_{4a}|_{k=p-p'}+\mathcal{M}_{4b}|_{k=p-p'}+\mathcal{M}_{4c}|_{k=p+p'+(\lambda_1-\lambda_2)P}.
\end{equation}

In the rest frame, we define $p_1'=(E_1',-\mathbf{p}'_1)$ and $p_2'=(E_2',\mathbf{p}'_2)$ to be the momenta of $\pi$ and $B_s^\ast$, respectively. According to the kinematics of two-body decay of the initial state in the rest frame, one has
\begin{equation}
\begin{split}
E_1'&=\frac{M^2-m_2^{'2}+m_1^{'2}}{2M},\quad\quad E_1'=\frac{M^2-m_1^{'2}+m_2^{'2}}{2M},\\
|\mathbf{p}'_1|&=|\mathbf{p}'_2|=\frac{\sqrt{[M^2-(m'_1+m'_2)^2][M^2-(m'_1-m'_2)^2]}}{2M},
\end{split}
\end{equation}
and
\begin{equation}
d\Gamma=\frac{1}{32\pi^2}|\mathcal{M}|^2\frac{|\mathbf{p}'|}{M^2}d\Omega,
\end{equation}
where $|\mathbf{p}'_1|$ and $|\mathbf{p}'_2|$ are the norm of the 3-momentum of the particles in the final states in the rest frame of the initial bound state and $\mathcal{M}$ is the Lorentz-invariant decay amplitude of the process.

As stated in Ref. \cite{Faessler:2008vc}, the couplings of $\pi^0$ to the $B^\ast B$ and $K^\ast K$ mesonic pairs contain two terms, i.e. the ``dominant'' coupling (proportional to $\cos\varepsilon$) and the ``suppressed'' coupling (proportional to $\sin\varepsilon$). This means that the first coupling survives in the isospin limit, while the second one vanishes.
\subsection{The radiative decays of $B_{s1}(5778)\rightarrow B_s^{0}\gamma$ and $B_{s1}\rightarrow B_s^{*0}\gamma$}

To estimate the radiative decays of the $B_{s1}(5778)$, we need additional effective Lagrangians related to the photon field, $A_\mu$, which are \cite{Chen:2010re}
\begin{equation}
  \begin{split}
\mathcal{L}_{B^{\ast}B^{\ast}\gamma}=&ieA_\mu(g^{\alpha\beta}B^\ast_\alpha\stackrel{\leftrightarrow}{\partial^{\mu}}B^\ast_\beta+g^{\mu\beta}B^\ast_\alpha\partial^{\alpha}B^\ast_\beta-
                                     g^{\mu\alpha}\partial^{\beta}B^\ast_\alpha B^\ast_\beta),\\
\mathcal{L}_{B^{\ast}B\gamma}=&\frac14g_{B^{\ast}B\gamma}e\varepsilon^{\mu\nu\alpha\beta}F_{\mu\nu}(\partial_{\alpha}B^\ast_\beta-\partial_{\beta}B^\ast_\alpha)B,\\
\mathcal{L}_{K^{\ast}K\gamma}=&\frac14g_{K^{\ast}K\gamma}e\varepsilon^{\mu\nu\alpha\beta}F_{\mu\nu}(\partial_{\alpha}K^\ast_\beta-\partial_{\beta}K^\ast_\alpha)K,\\
\mathcal{L}_{KK\gamma}=&ieA_\mu K\stackrel{\leftrightarrow}{\partial^{\mu}}K,
  \end{split}
\end{equation}
where the strength tensor are defined as $F_{\mu\nu}=\partial_\mu A_\nu-\partial_\nu A_\mu$.

\begin{figure}[ht]
\centering
    \rotatebox{0}{\includegraphics*[width=0.50\textwidth]{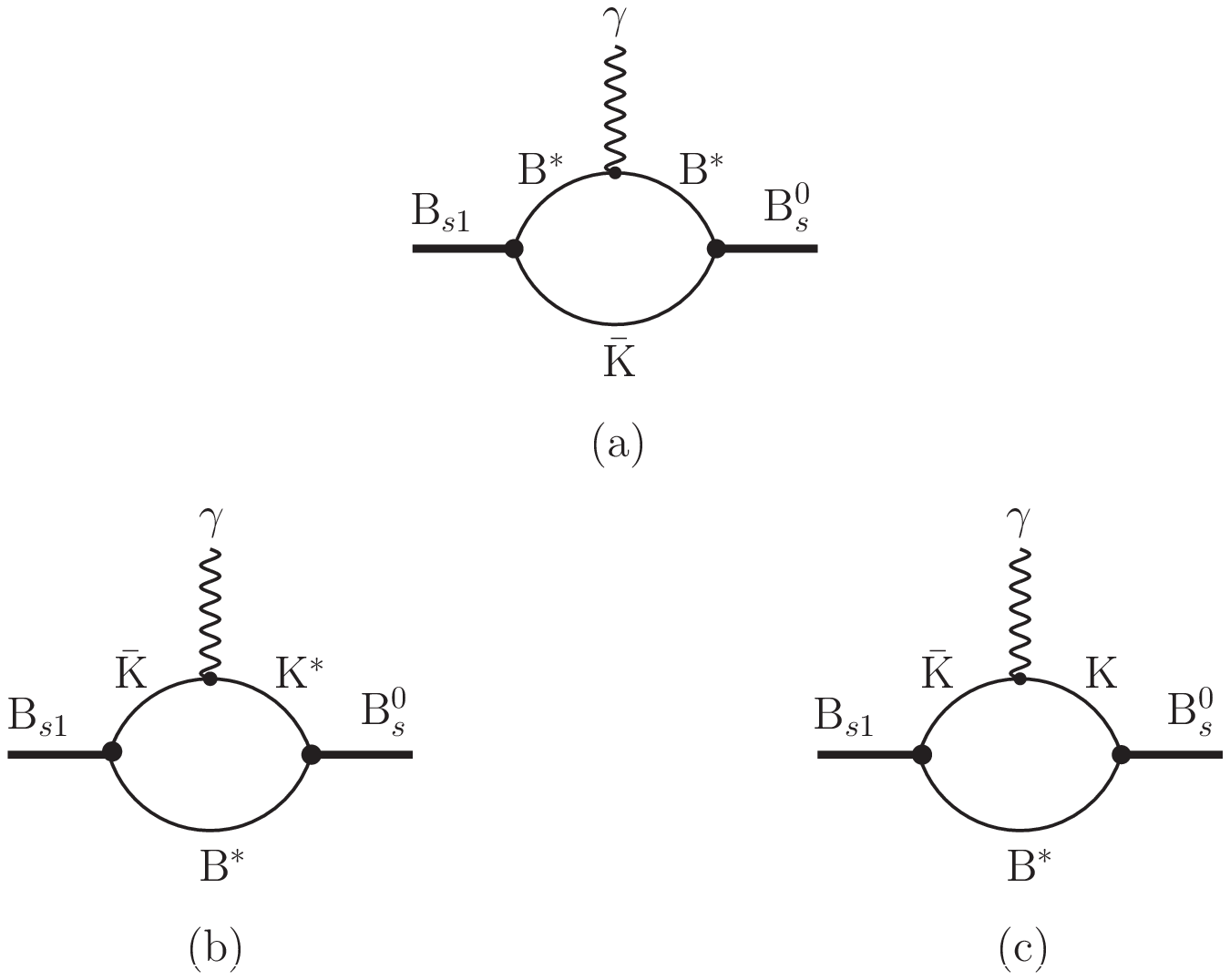}}
    \caption{Diagrams contributing to the $B_{s1}\rightarrow B_s^{0}\gamma$ decay.}
  \label{decay2}
\end{figure}

For the radiative decays $B_{s1}(5778)\rightarrow B_s^{0}\gamma$ and $B_{s1}(5778)\rightarrow B_s^{*0}\gamma$, we define $p_1'$ to be the momentum of $\gamma$, $p_2'$ to be the momentum of $B_s^{0}$ or $B_s^{*0}$.

In the radiative decay of $B_{s1}(5778)\rightarrow B_s^{0}\gamma$, the photon can be emitted from the bottom meson or the kaon. The diagrams are listed in Fig. \ref{decay2} and the corresponding amplitudes are
\begin{equation}
\begin{split}
\mathcal{M}_{5a}&=eg_{B^\ast BP}\epsilon_{\gamma}^\nu \int\frac{d^4p}{(2\pi)^4}p_{2\beta}[g_{\alpha\mu}(p_1-k)_\nu+g_{\nu\mu}p_{1\alpha}-g_{\alpha\nu}k_\mu]\Delta_{B^\ast}^{\alpha\beta}(k,m_{B^\ast})\mathcal{F}^2(k)\chi_P^\mu(p),\\
\mathcal{M}_{5b}&=\frac{e}{2}f_{B^\ast BV}g_{K^\ast K\gamma}\epsilon_{\rho\alpha\sigma\mu}\epsilon_{\kappa\lambda\delta\tau}\epsilon_{\gamma\nu} \int\frac{d^4p}{(2\pi)^4} k^\rho(p_1+p'_2)^\sigma(p_1^{'\kappa}g^{\nu\lambda}-{p_1'}^{\lambda}g^{\mu\kappa})\\
&\times(k^\delta g^{\beta\tau}-k^\tau g^{\beta\delta})\Delta_{K^\ast\beta}^\alpha(k,m_{K^\ast})\mathcal{F}^2(k)\chi_{P}^\mu(p),\\
\mathcal{M}_{5c}&=-eg_{B^\ast B_sP}\epsilon_\gamma^\nu \int\frac{d^4p}{(2\pi)^4}k_\mu(k+p_2)_\nu\Delta_K(k,m_K)\mathcal{F}^2(k)\chi_P^\nu(p),\\
\end{split}
\end{equation}
where $\epsilon_{\gamma}$ is the polarization of the photon.

The total amplitude for $B_{s1}(5778)\rightarrow B_s^{0}\gamma$ is then
\begin{equation}
\mathcal{M}_{B_{s1}(5778)\rightarrow B_s^{0}\gamma}=\mathcal{M}_{5a}|_{k=p-p'}+\mathcal{M}_{5b}|_{k=p+p'+(\lambda_1-\lambda_2)P}+\mathcal{M}_{5c}|_{k=p+p'+(\lambda_1-\lambda_2)P}.
\end{equation}

\begin{figure}[ht]
\centering
    \rotatebox{0}{\includegraphics*[width=0.50\textwidth]{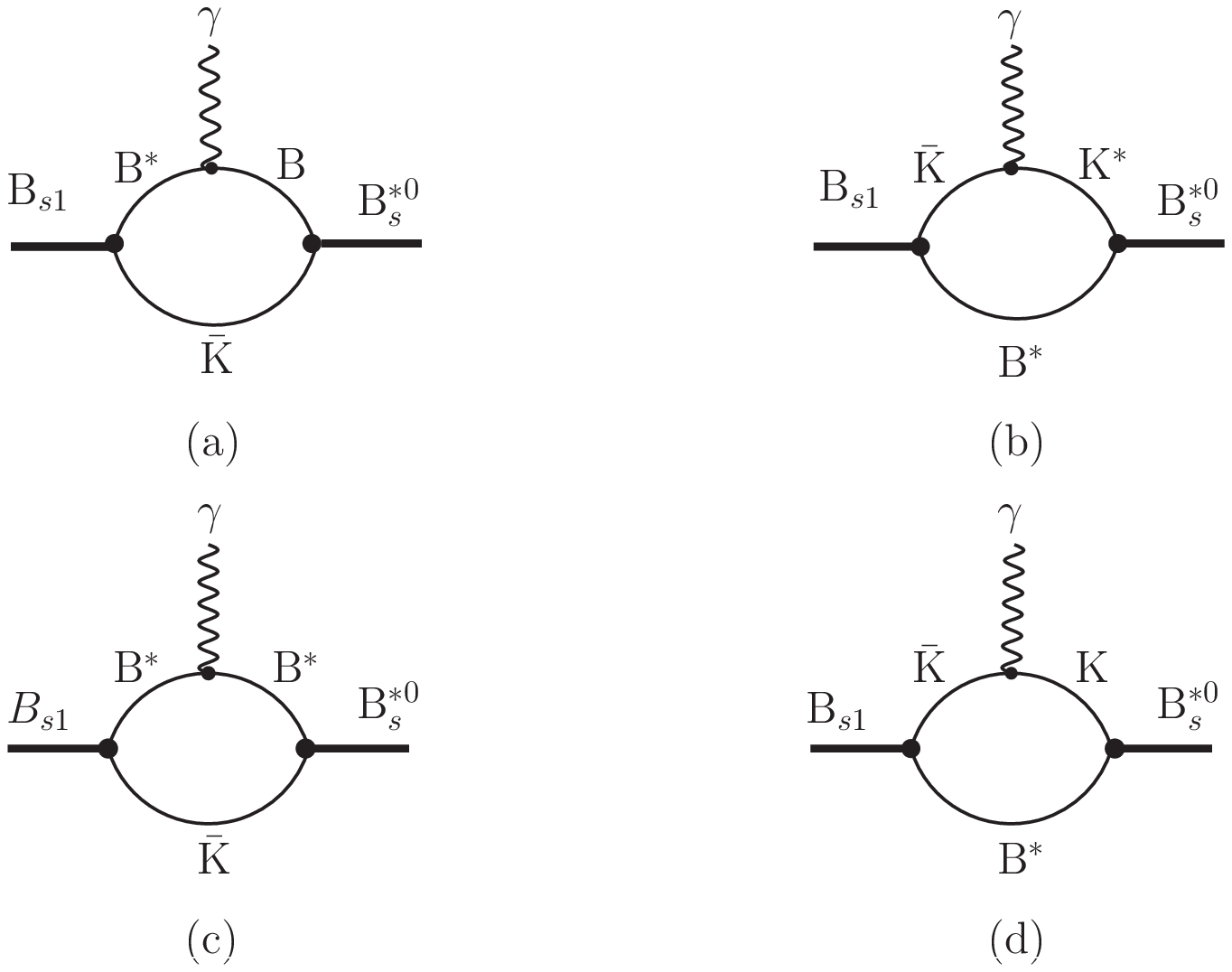}}
    \caption{Diagrams contributing to the $B_{s1}\rightarrow B_s^{*0}\gamma$ decay.}
  \label{decay3}
\end{figure}

For the $B_{s1}\rightarrow B_s^{*0}\gamma$ process, it is indicated in Ref. \cite{Faessler:2008vc} that the dominant contributions to $B_{s1}\rightarrow B_s^{*0}\gamma$ come from Fig.\ref{decay3}(a) and Fig.\ref{decay3}(b) which are gauge invariant. These amplitudes are almost one order bigger than those of Fig.\ref{decay3}(c) and Fig.\ref{decay3}(d).

The amplitudes for Fig.\ref{decay3}(a) and Fig.\ref{decay3}(b) are
\begin{equation}
\begin{split}
\mathcal{M}_{6a}&=\frac{e}{4}g_{B^\ast B\gamma}g_{B_s^\ast BP}\epsilon^{\tau\delta\sigma\rho}\epsilon_\gamma^\nu\epsilon_{B_s^\ast}^\lambda \int\frac{d^4p}{(2\pi)^4}p_{2\lambda}(p'_{1\tau}g_{\nu\delta}-p'_{1\delta}g_{\nu\tau})(p_{1\sigma}g_{\mu\rho}-p_{1\rho}g_{\sigma\mu})\Delta_B(k,m_B)\mathcal{F}^2(k)\chi_P^\mu(p),\\
\mathcal{M}_{6b}&=\frac{e}{4}g_{K^\ast K\gamma}\epsilon_{\rho\delta\tau\sigma}\epsilon_\gamma^\nu\epsilon_{B_s^\ast}^\lambda\int\frac{d^4p}{(2\pi)^4}(p'_{1\rho}g_{\nu\delta}-p'_{1\delta}g_{\nu\rho})(k_{\tau}g_{\sigma\beta}-k_{\sigma}g_{\beta\tau})\\
&\times[g_{B_s^\ast B^\ast V}(p_1+p'_2)_\alpha g_{\mu\lambda}+4f_{B_s^\ast B^\ast V}(k_\lambda g_{\alpha\mu}-k_\mu g_{\alpha\lambda})]\Delta_{K^\ast}^{\alpha\beta}(k,m_{K^\ast})\mathcal{F}^2(k)\chi_P^\mu(p).
\end{split}
\end{equation}
where $\epsilon_{B_s^\ast}$ is the polarization of $B_s^\ast$ meson.

The total amplitude for $B_{s1}\rightarrow B_s^{*0}\gamma$ is then
\begin{equation}
\mathcal{M}_{B_{s1}\rightarrow B_s^{*0}\gamma}=\mathcal{M}_{6a}|_{k=p-p'}+\mathcal{M}_{6b}|_{k=p+p'+(\lambda_1-\lambda_2)P}.
\end{equation}

\subsection{Numerical results}

In the calculation, we use the following input parameters \cite{Tanabashi:2018oca}: $m_{\pi^0}$ = 134.977 MeV, $m_{K^\pm}$ = 493.677 MeV, $m_{K^0}$ = 497.611 MeV, $m_{K^{\ast\pm}}$ = 891.76 MeV, $m_{K^{\ast0}}$ = 895.55 MeV, $m_{B^\pm}$ = 5279.32 MeV, $m_{B^0}$ = 5279.63 MeV, $m_{B^{\ast0}}$ = 5324.65 MeV, $m_{B_s^{0}}$ = 5366.89 MeV, $m_{B_s^{\ast0}}$ = 5415.4 MeV. We apply the normalized numerical solutions of the BS equation and the corresponding cutoff $\Lambda$ for different form factors to the decays calculation, and obtain the following predictions for the decay widths:
\begin{equation}
\begin{split}\label{decay-widths1}
\Gamma_{B_{s1}\rightarrow B_s^\ast\pi} &= 27.5 \,\mathrm{KeV},\,\quad\text{when}\, \Lambda_M=1541\, \mathrm{MeV},\\
\Gamma_{B_{s1}\rightarrow B_s^\ast\pi} &= 34.7\,\mathrm{KeV},\,\quad\text{when}\, \Lambda_D=2210\, \mathrm{MeV},\\
\Gamma_{B_{s1}\rightarrow B_s^\ast\pi} &= 39.2 \,\mathrm{KeV},\,\quad\text{when}\, \Lambda_E=1591\, \mathrm{MeV},\\
\end{split}
\end{equation}

\begin{equation}
\begin{split}\label{decay-widths2}
\Gamma_{B_{s1}\rightarrow B_s\gamma} &= 45.2 \,\mathrm{KeV},\,\quad\text{when}\, \Lambda_M=1541\, \mathrm{MeV},\\
\Gamma_{B_{s1}\rightarrow B_s\gamma} &= 64.3\,\mathrm{KeV},\,\quad\text{when}\, \Lambda_D=2210\, \mathrm{MeV},\\
\Gamma_{B_{s1}\rightarrow B_s\gamma} &= 79.8\,\mathrm{KeV},\,\quad\text{when}\, \Lambda_E=1591\, \mathrm{MeV},\\
\end{split}
\end{equation}

\begin{equation}
\begin{split}\label{decay-widths3}
\Gamma_{B_{s1}\rightarrow B_s^{\ast}\gamma} &= 0.4 \,\mathrm{KeV},\,\quad\text{when}\, \Lambda_M=1541\, \mathrm{MeV},\\
\Gamma_{B_{s1}\rightarrow B_s^{\ast}\gamma} &= 1.9\,\mathrm{KeV},\,\quad\text{when}\, \Lambda_D=2210\, \mathrm{MeV},\\
\Gamma_{B_{s1}\rightarrow B_s^{\ast}\gamma} &= 2.6\,\mathrm{KeV},\,\quad\text{when}\, \Lambda_E=1591\, \mathrm{MeV},\\
\end{split}
\end{equation}
where the values 1541, 2210, 1591 MeV correspond to the monopole, dipole, and exponential form factors, respectively.

In comparison, we also display the predictions for $B_{s1}\rightarrow B_s^\ast\pi$, $B_{s1}\rightarrow B_s\gamma$, and $B_{s1}\rightarrow B_s^{\ast}\gamma$ decay widths from other theoretical approaches in Table \ref{decay-width}. Ref. \cite{Bardeen:2003kt} is based on the chiral symmetry in the heavy-light meson system. In Ref. \cite{Guo:2006rp} $B_{s1}$ is considered as a $B^\ast\bar{K}$ bound state in heavy chiral unitary approach. The strong and radiative decays are calculated using light-cone QCD sum rules \cite{Wang:2008ny,Wang:2008wz}. The radiative decay widths of $B_{s1}$ are studied in a pure $b\bar{q}$ structure, and a mixed one, ($b\bar{q}+bq\bar{q}\bar{q}$) in Ref. \cite{Vijande:2007ke}. Also assuming $B_{s1}$ as a $B^\ast \bar{K}$  hadronic molecule, the authors of Ref. \cite{Faessler:2008vc} and Ref. \cite{Cleven:2014oka} analyzed its strong and radiative decay widths by using a phenomenological Lagrangian approach.

In Refs. \cite{Faessler:2008vc,Cleven:2014oka} and our work, $B_{s1}$ is all regarded as a $B^\ast\bar{K}$ molecular state. In Ref. \cite{Cleven:2014oka}, the mass of the $B^\ast K$ bound states is obtained as 5671$\pm$45 MeV by using the chiral perturbation theory approach, which is around 100 MeV lower than those in our work and Ref. \cite{Faessler:2008vc} (we all taken the mass of $B_{s1}$ from Ref. \cite{Guo:2006rp}, in which the authors also predicted the mass of $B_{s1}$ by using the chiral perturbation theory approach.), and it is found that the mass of $B_{s1}$ has a great influence on the decay width. For example, when the mass of $B_{s0}^\ast$ varies from  5625$\pm$45 MeV to 5725 MeV, the decay width of $B_{s0}^\ast\rightarrow B_s\pi^0$ varies from 0.8$\pm$0.8 keV to 73 keV. In Ref. \cite{Faessler:2008vc}, the $B_{s1}\rightarrow B_s^\ast\pi^0$, $B_{s1}\rightarrow B_s\gamma$ and $B_{s1}\rightarrow B_s^\ast\gamma$ decay widthes are evaluated considering the $K^\ast$, $K$ and $K^\ast$ meson exchange diagram contributions, respectively. In our calculation, we consider all the possible meson exchange diagrams, i.e. the $B$, $B^\ast$ and $K^\ast$ mesons exchange diagrams in $B_{s1}\rightarrow B_s^\ast\pi^0$ decay, $B^\ast$, $K^\ast$ and $K$ mesons exchange in $B_{s1}\rightarrow B_s\gamma$ decay, and $B$ and $K^\ast$ mesons exchange diagrams in $B_{s1}\rightarrow B_s^\ast\gamma$ decay, respectively.
Another difference is the value of the coupling constant $g_{B^\ast B_sK}$, we have applied the heavy-quark symmetry and chiral symmetry as in Ref. \cite{Casalbuoni:1996pg} and adopt $g_{B^\ast B_sK}=\frac{2g}{f_\pi}\sqrt{m_{B^\ast}m_{B_s}}=35.35$, while in Ref.\cite{Faessler:2008vc}, $g_{B^\ast B_sK}$ = 5.70 is used which is from QCD sum rules. This has a great effect on the calculation of the decay widths.

\begin{table}[tb]
\renewcommand{\arraystretch}{1.2}
\centering
\caption{
The decay widths (in KeV) of $B_{s1}\rightarrow B_s^\ast\pi$, $B_{s1}\rightarrow B_s\gamma$, and $B_{s1}\rightarrow B_s^{\ast}\gamma$ in various theoretical approaches.}
\begin{tabular*}{\textwidth}{@{\extracolsep{\fill}}cccccc}
\hline
\hline
                            & $\Gamma({B_{s1}\rightarrow B_s^\ast\pi})$ &                           & $\Gamma({B_{s1}\rightarrow B_s\gamma})$ &                           & $\Gamma({B_{s1}\rightarrow B_s^{\ast}\gamma})$ \\
\hline
Ref. \cite{Bardeen:2003kt}  &21.5                                       &Ref. \cite{Bardeen:2003kt} &39.1                                     &Ref. \cite{Bardeen:2003kt} & 56.9                                           \\
Ref. \cite{Guo:2006rp}      &10.36                                      &Ref. \cite{Wang:2008wz}    &3.2-15.8                                 &Ref. \cite{Wang:2008wz}    & 0.3-6.1                                        \\
Ref. \cite{Wang:2008ny}     &5.3-20.7                                   &Ref. \cite{Vijande:2007ke} &106.5(60.7)                              &Ref. \cite{Vijande:2007ke} & 75.6(0.6)                                      \\
Ref. \cite{Faessler:2008vc} &57.0-94.0                                  &Ref. \cite{Faessler:2008vc}&2.01-2.67                                &Ref. \cite{Faessler:2008vc}& 0.04-0.18                                      \\
Ref. \cite{Cleven:2014oka}  &1.8$\pm$1.8                                &Ref. \cite{Cleven:2014oka} &4.1$\pm$10.9                             &Ref. \cite{Cleven:2014oka} & 46.9$\pm$33.6                                    \\
\hline
\hline
\end{tabular*}\label{decay-width}
\end{table}

\section{Summary}
\label{sum}
In this work, we studied the bottom-strange meson $B_{s1}(5778)$ with the hadronic molecule interpretation, i.e. regarding it as a bound state of $B^\ast\bar{K}$ meson in the BS equation approach. In our model, we applied the ladder and instantaneous approximations to obtain the kernel containing one particle-exchange diagrams and introduced three different form factors (monopole form factor, dipole form factor, and exponential form factor), since the constituent particles and the exchanged particles in the $B^\ast\bar{K}$ system are not pointlike. The cutoff $\Lambda$ which was introduced in the form factors reflects the effects of the structure of interacting particles. Since $\Lambda$ is controlled by nonperturbative QCD and cantnot be determined at accurately, we let it vary in a reasonable range within which we try to find possible bound states of the $B^\ast\bar{K}$ system.

From our calculations, we found that there exist isoscalar bound states of the $B^\ast\bar{K}$ system when $\Lambda$ (and $E_b$ correspondingly) varies in a range. The bound state of $B^\ast\bar{K}$ can be assigned to the $B_{s1}(5778)$ state when the cutoff $\Lambda$ =1541 MeV, 2210 MeV, and 1591 MeV for the monopole, dipole, and exponential form factors, respectively. With the obtained numerical results for the normalized BS wave function, we also calculated the decay widths of the decay $B_{s1}(5778)\rightarrow B_s^\ast\pi$ including the $\eta-\pi^0$ mixing effect and the radiative decays $B_{s1}(5778)\rightarrow B_s\gamma$ and $B_{s1}(5778)\rightarrow B_s^{\ast}\gamma$. We predict that the decay widths are 27.5 KeV, 34.7 KeV, and 39.2 KeV for $B_{s1}(5778)\rightarrow B_s^\ast\pi$, 45.2 KeV, 64.3 KeV, and 79.8 KeV for $B_{s1}(5778)\rightarrow B_s\gamma$, and 0.4 KeV, 1.9 KeV, and 2.6 KeV for $B_{s1}(5778)\rightarrow B_s^{\ast}\gamma$ for the monopole, dipole and exponential form factors, respectively. We expect forthcoming experimental measurements to test our model for the state $B_{s1}(5778)$ as a $B^\ast\bar{K}$ molecule.

\acknowledgments
One of the authors (Z.-Y. W.) is very grateful to Professor Zhen-Hua Zhang, Professor Qin Chang and Professor Jia-Jun Wu for valuable discussions. Qi-Xin Yu acknowledges the support from the China Scholarship
Council. This work was supported by National Natural Science Foundation of China (Projects No. 11775024, No.11575023 and No.11605150) and K.C.Wong Magna Fund in Ningbo University.

\end{document}